Title
Optical and Strain Stabilization of Point Defects in Silicon Carbide

Authors:
Jonathan R. Dietz, Evelyn L. Hu

Correspondence: jdietz@g.harvard.edu

John A. Paulson School of Engineering and Applied Sciences, Harvard University, Cambridge, Massachusetts 02138, USA

*Abstract*

The photoluminescence and spin properties of ensembles of color centers in silicon carbide are enhanced by fabricating optically isolated slab waveguide structures and carefully controlling annealing and cooling conditions. We find that the photoluminescence signal of an ensemble of implanted defects is enhanced in slab waveguides by an order of magnitude over identically implanted bulk defects. The slab waveguide-enhanced photoluminescence of several defect species is used to study recombination and diffusion in the presence of thermal annealing with both rapid quench cooling and a longer return to ambient conditions. The confined mechanical geometry of the thin film is exploited to measure the spin-strain coupling of the negatively charged silicon monovacancy. The methods in this work can be used to exercise greater control on near-surface emitters in silicon carbide and better understand and control the effects of strain on spin measurements of silicon carbide based color centers.

*Body*

Silicon carbide is a wide-band gap semiconductor with many color centers that emit from the visible to the infrared; many of the centers have paramagnetic resonances useful for the construction of quantum memories, single photon sources, magnetometers, and electrometers, to name a few.[1–5] While both intrinsic and extrinsic color centers benefit from silicon carbide's wealth of manufacturing processes and accessibility as a wafer scale semiconductor platform, numerous challenges remain in realizing high quality device applications. The performance of the color centers is often subject to limitations in brightness and indistinguishability. Beyond issues of extracting light from atomic scale emitters within high index of refraction materials, there are other reasons leading to the relatively low detected photon emission rates of the negatively charged silicon vacancy $V_{Si}^-$ and neutral divacancy $V_{Si}V_C^0$. There is a fast non-radiative decay into the intersystem crossing states of the defect system as compared to its spin conserving radiative decay.[6] Previous work address this by coupling emitters to photonic crystal cavities, resulting in a three-fold decrease in the spontaneous emission lifetime, producing an increased luminescence output.[7] In addition, *charge state conversion* of a defect state can quench an optically bright color center.[8] Finally, the dipole moment of the center's optical transition may be poorly aligned with its optical excitation. In this study we demonstrate that slab waveguides alone can significantly enhance the optical and spin properties of surface ensembles in silicon carbide. We leverage this enhancement to study the recombination of SiC defect centers under different annealing-and-cooling processes, noting the changes in optical and spin signatures of the ensemble.

The formation of SiC defect centers with controlled spatial placement and defect type is critical for their implementation within quantum information systems, strongly depending on the means of defect formation and annealing. Forming centers in pure, epitaxially grown material requires irradiation with energetic photons, electrons, or ions which introduces residual damage in the form of interstitials, secondary defects, surface states, and stress.[9–11] These effects can produce broadened zero phonon lines, spin resonances, and reduced signal to noise in spin read-out. Post implantation treatment mitigates some of these effects, recombining unwanted defect species and stabilizing desired defect species.[12–14] While initial studies have been performed to study the critical temperatures of formation for specific defects, there is need for comprehensive experimental work studying defects' formation, recombination, and diffusion. Fortunately, many of silicon carbide's intrinsic defects have bright optical transitions and spin transitions, useful in studying the relative defect populations and assessing overall material strain. In this work, we study the optical signal of the bright intrinsic defects in silicon carbide: carbon anti-site vacancy ($C_{Si}V_C^{2+}$), silicon mono-vacancy ($V_{Si}^-$), and divacancy ($V_{Si}V_C^0$) and the optically detected magnetic resonance (ODMR) of the k-site $V_{Si}^-$ or V2 line. Starting from c-axis grown material, we first study the way that undercut material produces slab waveguide modes that efficiently scatter pump photons into the typically inefficiently excited V1 and V2 lines and the consequent enhancement of the magnetic resonance of the V2 line. Then, using our enhanced ability to identify the constituent defects, we intentionally modify the ensemble, through annealing in air at 600C to produce $C_{Si}V_C^{2+}$ and at 850C to produce $V_{Si}V_C^0$, followed by either slow cooling or rapid quenching. We monitor the relative population of each defect species after each annealing and cooling process by photoluminescence (PL) measurements and monitor the thin film stress via optically detected magnetic resonance (ODMR) and subsequent focused ion beam stress-strain release.

The two samples used in this study were fabricated from epitaxially grown 4H SiC (Norstel AB). Starting from a bulk, n-type ($10^{19}$ cm$^{-3}$) wafer of 4H-SiC, layers of p-type ($10^{18}$ cm$^{-3}$), i-type, and p-type ($10^{18}$ cm$^{-3}$) material were epitaxially grown in layer thicknesses of 100 nm, 200 nm, and 100 nm, respectively. Electron-beam patterning formed patterns of rectangular nanobeam waveguides, approximately 200 nm thick by 200 nm wide, and 10 microns in length. The nanobeams are suspended through an undercut process that employs a photo-electrochemical (PEC) process, as described in previous work.[7,15,16] The PEC process also results in areas of undercut material that form 2um wide and 200-400nm thick slab waveguides, that extend for distances of 20 microns. The samples are then implanted at a dose of 1e12 $C^{12}$ ions, and energy of 70 keV, at 7 degrees off of the c-axis to form an ensemble of color centers.

Thermal annealing was carried out on the samples, using a quartz tube reactor in air, at two temperatures. An initial anneal was performed at 600° C for one hour. The goal of this anneal was to allow diffusion of those SiC defects with higher diffusion constants, and to form $C_{Si}V_C$.[2] A subsequent anneal was carried out at 850° C for 1 hour; these are the conditions that have been reported to form $V_{Si}V_C$.[17] After each high temperature step, return to room temperature was accomplished either through slow cooling over 6 hours or by rapid quenching through immersion in room temperature water.

The PL and V2 ODMR of the ensemble is measured after each thermal annealing and subsequent cooling step. All PL and ODMR measurements are performed in a continuous flow cryostat (Janis ST-500) at 77K. Excitation (collection) light is delivered (collected) with an NA 0.6 objective, and then filtered through a 50um pinhole. The PL of $C_{Si}V_C$ and $V_{Si}$ is measured with 300uW of 632.8nm from a helium-neon laser. The PL of $V_{Si}V_C$ and k-site $V_{Si}$ centers is measured with 2mW of 865nm laser light.

ODMR is measured, using a home built confocal ODMR spectrometer equipped with a single photon avalanche detector with 100uW of 865nm excitation light and luminescence filtered with 900nm long-pass and 1000nm short-pass filters to eliminate the phonon sideband of h-site $V_{Si}^-$ and $VV^0$ PL. Radio-frequency excitation is supplied from an RF signal generator (Stanford Research System SR384) and then amplified to 30dBm (Minicircuits ZHL-2-S+) which is directed to the sample by a closely placed 50um wire.

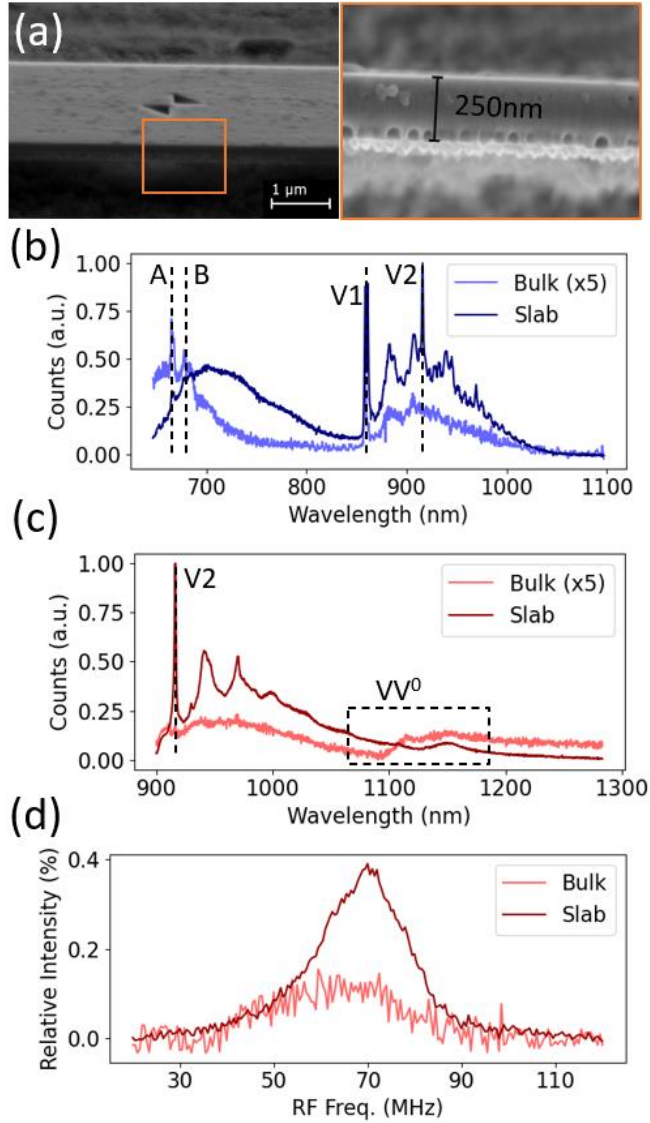

Fig. 1. Slab waveguide enhanced properties of intrinsic defects of silicon carbide. (a) Field-emission scanning electron microscope image of suspended waveguide structure. Inset: image demonstrating roughened morphology. (b) Photoluminescence of as-implanted intrinsic color centers excited by 632nm light. (c) Photoluminescence of k-site $V_{Si}^-$ (V2) center and weak divacancy in bulk and slab waveguides excited by 865nm light and its (d) corresponding optically detected magnetic resonance.

Implantation into SiC can gives rise to several defect species: in addition to the $V_{Si}$, there are $V_C$'s (carbon vacancies of different charge states), Si and C interstitials, $V_{Si}V_C$ (silicon-carbon divacancies), and Si and C anti-site defects. $C_{Si}V_C$ has an important impact on the number of $V_{Si}$'s observed, as annealing allows carbon atoms to convert to or from $V_{Si}$ in an equilibrium reaction[13,14]:

$$V_{Si} \leftrightarrow C_{Si}V_C$$

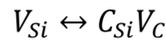

To gain insight into this conversion process at different annealing conditions, our data will track the emission of both the $V_{Si}^-$, and the carbon anti-site vacancy pair ($C_{Si}V_C^{2+}$). Carbon anti-site vacancies have two sets of visible ZPLs corresponding to its inequivalent lattice configurations. ZPLs at 640-670nm are denoted as A-lines and ZPLs at 670-680nm are denoted as B-lines.[18]

Figures 1(b) and 1(c) compare the PL of the unannealed defect ensemble in the bulk sample, to the PL in the slab waveguide region. Figure 1(b) delineates the spectral region from ~ 600 nm – 1100 nm, while Figure 1(c) delineates the region from 900-1300 nm, where we look for $V_{Si}^-$ and possible $V_{Si}V_C^0$ emission. Although there is some indication of $C_{Si}V_C^{2+}$ emission in the bulk material, broad background emission in the slab waveguide, peaking at 700-750nm, obscures $C_{Si}V_C^{2+}$ emission. Figures 1(b) and 1(c) reveal an order of magnitude higher output intensity for silicon vacancy emission and strong V1 and V2 zero-phonon lines (ZPL). These ZPLs do not appear in bulk spectra, only V1' appears. We believe that the enhanced luminescence results from the slab waveguide geometry, a result of multiple reflections at the top and bottom surfaces of the slab. In addition, the strong appearance of both V1 and V2 ZPLs in the slab waveguide suggests a larger intermixing of TM-mode field into the waveguide, enhancing coupling to the V2 transition. V1 and V2 have dipoles aligned nearly orthogonally to the basal plane, coupling poorly to the basal plane-oriented excitation in the bulk. The rough under-surface of the slab waveguide, as shown in scanning electron micrograph images, can result in mixing between the TE character of the laser excitation and the mostly TM character of the V1 and V2 dipole moments.

The ODMR signal of the ensemble is also significantly improved in the undercut material, compared to the bulk. As shown in Figure 1(d), for measurements taken with no applied magnetic field, the two degenerate spin transitions have broad, overlapping 20MHz linewidths. In the slab waveguide the magnetic resonance signature is twice as strong as in the bulk material, owing to the enhancement of the V2 dipole transition from slab waveguide confinement. When fitted with a bi-Lorentzian, there is a small splitting of 10MHz between the normally Kramer-degenerate $|-1/2\rangle \leftrightarrow |-3/2\rangle$ and $|1/2\rangle \leftrightarrow |3/2\rangle$ spin subspaces. We attribute this splitting to residual strain in the material that has developed over the course of fabrication and ion implantation.

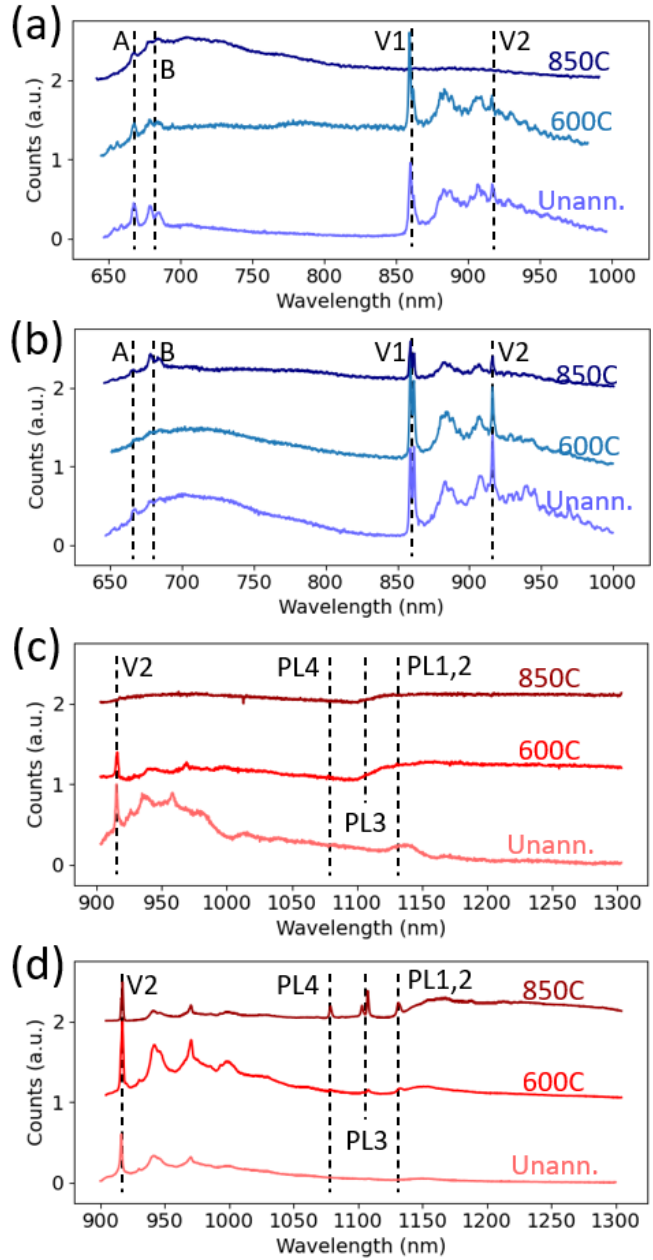

Fig. 2. Evolution of slab ensemble emission over the course of annealing with and without quench cooling. Each temperature condition is offset by 1 arbitrary unit. (a,b) Evolution of 632nm excited carbon anti-site and silicon monovacancy in the case of (a) slow cooling and (b) quench cooling. (c,d) Evolution of 865nm excited silicon monovacancy and divacancy emission in the case of (c) slow cooling and (d) quench cooling. Quench cooling is believed to limit conversion of $V_{Si}$ into $C_{Si}V_C$.

Thermal annealing of an implanted sample can remove residual unwanted material defects, and promote defect recombination. In general, $V_{Si}$ and $C_{Si}V_C$ ZPLs are stronger and better defined in slab samples as compared to bulk samples, so the data shown in Figure 2 refer to only slab waveguide samples. The improved optical read-out of the slab waveguide geometry allows us to monitor defects throughout annealing processes where defect diffusion and recombination may lower the density and hence detectability of the defects.

In Figures 2(a,c) the materials are annealed and then allowed to "naturally" return to room temperature. While the 600C annealing step increases the $V_{Si}$ fluorescence, further annealing at 850C results in a substantially diminished signal. The $C_{Si}V_C$ ZPL emission, present in the as-implanted material is no longer discernible after the 600C, and 850C annealing steps, masked within the broad 700nm luminescence peak. This trend, the diminution of the $V_{Si}$ with increased annealing temperature, is also observed in the spectrum of Figure 2(c), where the 850C anneal results in an essentially featureless spectrum.

Figures 2(b) and 2(d) display some dramatically different features when the annealing steps are followed by a "quench cooling" process to return the samples to room temperature. Previous work suggested that the conversion of $V_{Si}$ to $C_{Si}V_C$ through thermal annealing can be limited by rapidly cooling the sample, thus maintaining a relatively high density of $V_{Si}$ as compared to a sample that is slowly cooled.[12] Indeed, our results are consistent with those earlier findings. After quench-cooling from 600C, there is a partial reduction of the broad band centered at 700nm, and no anti-site luminescence appears. The divacancy band is strengthened and several ZPLs appear. After a further anneal at 850C and quench-cooling, the broad band at 700 nm is further suppressed, and the anti-site PL transition is evident. A decrease in $V_{Si}$ is accompanied by an increase in both the divacancy and anti-site emissions, with the additional appearance of all four $VV^0$ ZPLs. Finally, the spectrum directly in the vicinity of the V2 line shows minimal background fluorescence, indicating the suppression of PL that would decrease the ODMR signal-to-noise. Importantly, when compared to annealing without quenching, the ZPLs of all three defect species are stable and visible. The quench cooling may constrain the diffusion of the SiC defects, and thus constrain their recombination or aggregation.

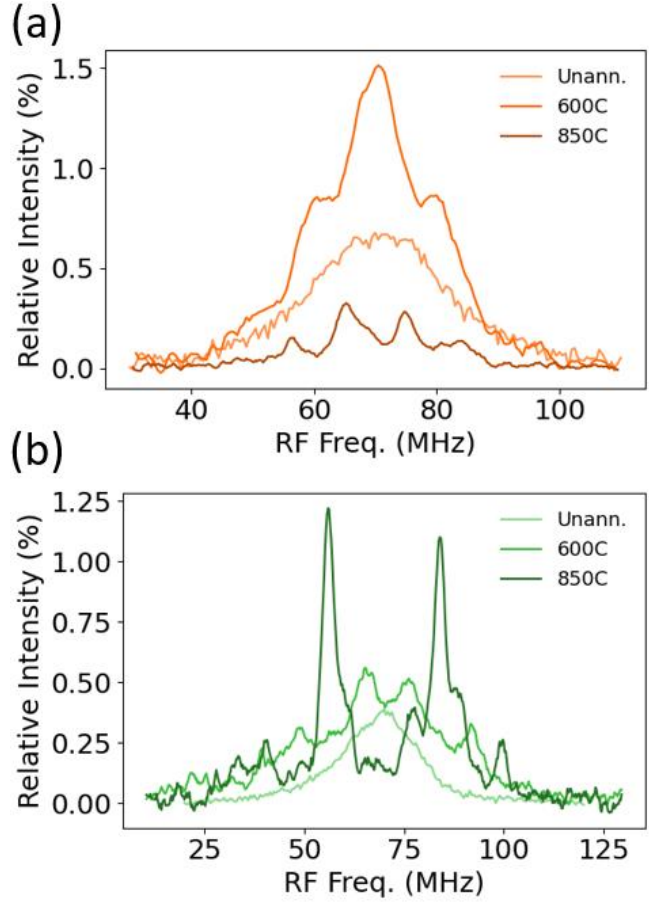

Fig. 3. Evolution of the V2 ODMR spectra (a) without quench cooling and (b) with quench cooling, demonstrating controlled relaxation/enhancement of strain. The longer time allowed in ordinary cooling allows for greater stress release in the sample, compared to the case with quench cooling.

|  | Quenched | | Unquenched | |
| --- | --- | --- | --- | --- |
|  | Splitting | Linewidth | Splitting | Linewidth |
| Unann. | 8.6±0.7MHz | 15.4±1.0MHz | 10.2±0.6MHz | 14.5±1.5MHz |
| 600C | 11.5±0.2MHz | 11.2±0.5MHz | 3.2±0.6MHz | 8.9±0.7MHz |
| 850C | 27.6±0.1MHz | 3.2±0.2MHz | 10.0±0.2MHz | 4.7±0.2MHz |

Table 1: Fitted results of strained ODMR traces shown in Fig.3.a and Fig.3.b.

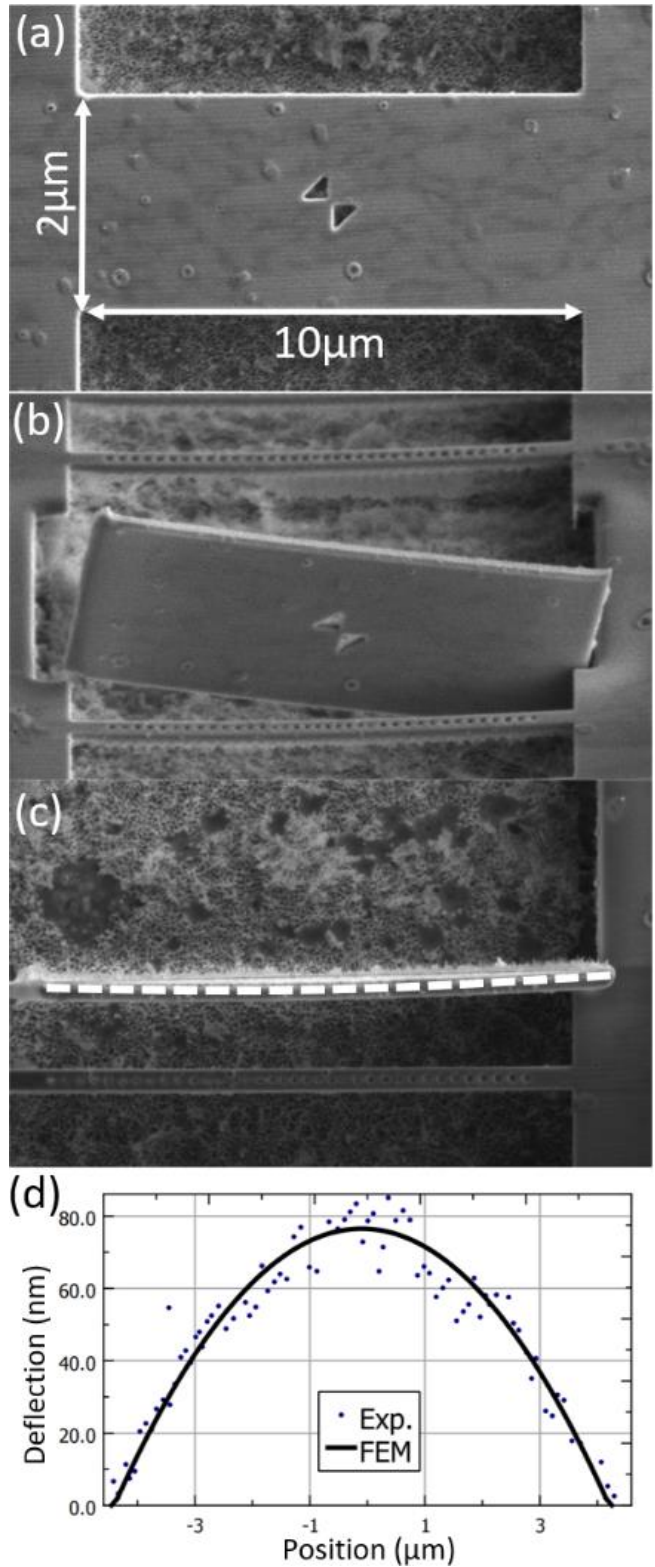

Fig. 4. Direct strain measurement through Focused Ion Beam (FIB) release. (a) Starting from a fully undercut slab waveguide (b) the waveguide is released by milling the edges to release the structure from its support and flipped upward with a micromanipulator. (c) Once flipped up, the stress-strain deflection (emphasized by a dotted line) of the released film can be measured directly to (d) infer the quench-cooling mediated stress concentration via finite element model analysis.

How do the ODMR traces track the various annealing processes? In the case of anneal with slow cooling, the initial 600C anneal improves the V2 ODMR signal, as shown in Figure 3(b), increasing the intensity of the resonance. The linewidth of the transition also narrows slightly, and the strain splitting nearly disappears. There are also satellites to the central ODMR transitions inconsistent with a typical hyperfine interaction, as they are split from the zero field splitting by three times the strain splitting of the central lines and lack the symmetric structure of hyperfine interactions.[19] When annealed further to 850C, ODMR intensity decreases commensurately with the integrated PL (Fig. 2c). Surprisingly the linewidth of the ODMR continues to decrease, reaching a fitted width of 4.7MHz. Strikingly, strain splitting redevelops, and the satellite peaks persist, as summarized in Table 1. For the quench-cooled samples, 600C annealing again results in an ODMR signal with increased intensity; additional satellite lines appear in the ODMR spectrum, and the transition linewidth narrows. The ODMR also indicates a substantially greater amount of the material strain for the quench-cooled samples. After the 850C anneal followed by quench-cooling, the splitting in the ODMR is further split, while the linewidth further decreases, dropping to 3.2MHz, revealing previously unobserved hyperfine lines. Following thermal treatment, residual film strain is assessed by using a Focused Ion Beam (FIB) of gallium ions to release the waveguide structure and then measuring the film deflection via electron microscope as shown in Figure 4. We estimate the strain in the released slab using a finite element COMSOL simulation and the measured displacement. Using 4H-SiC literature values for silicon carbide's elastic modulus (500GPa) and Poisson ratio (0.2) we estimate the film strain to be on the order of 0.2%.[20]

The experiments in this paper underscore the important interplay of material geometries, and the importance of tracking both luminescence and spin properties of multiple color centers, as various processing conditions are applied. Undercut structures are the basis of optically isolated cavities that can selectively enhance color-center transitions. At a simpler level, the more limited volumes probed (compared to bulk samples), and their visible display of strain, allows us to better track processes of defect formation and recombination. By examining the relative intensity of three bright intrinsic defects, $V_{Si}$, $C_{Si}V_C$, and $V_{Si}V_C$, in the presence of annealing with or without a subsequent quench-cooling step, it is possible to observe several performance limiting changes. The first is the presence of background luminescence, which is shown to obscure the ZPLs of $C_{Si}V_C$ and also results in poor V2 ODMR contrast. Additionally, it is shown that this undesired emission is enhanced through annealing in air when not rapidly cooled. By contrast, when quench-cooling is employed, that undesired emission is suppressed, likely through more limited defect diffusion and recombination during cooling. Finally, our results support a premise that quench-cooling inhibits the recombination of $V_{Si}$, thus significantly enhancing its emission.

The understanding of material strain and its influence on ODMR signatures is critical for control of color centers for device applications, particularly in implanted and near surface emitters. Through careful thermal annealing it will be possible to mitigate some of these strain effects. Static strain splitting can be either mitigated through careful annealing to release strain or enhanced via rapid cooling. Annealing contributes to significantly narrowed ODMR lines, which can lead to improved sensitivity as a magnetometer or strain sensor.

Several elements remain to be understood. First and foremost is the identity of the luminescence peaking at 700nm. This emission band clearly affects the signal to noise of the $V_{Si}$ PL and ODMR contrast in the annealed ensemble. An increase in this emission is clearly associated with the quenching of all three of the studied defect species. Second is understanding the intrinsic stress in the SiC starting material. As a S=3/2 defect system, $V_{Si}$ should retain its Kramer's degeneracy under strain, whereas in our work the Kramer's degeneracy is clearly lifted, in contrast to theoretical prediction and experiment in 6H-SiC.[21,22]

Further limiting the volume of the undercut material to smaller dimensions, as applied to nanobeam cavities, with well-controlled annealing conditions may open the way towards defect recombination studies having greater precision.[23] Although the surface roughness of undercut structures may introduce unwanted surface states, periodically roughened surfaces may allow us to carry out a better coupling of external laser excitation to match color center dipoles. Finally, careful selection of annealing and thermal quenching conditions yields ensembles with a highly homogeneous strain bias, which could lead to more sensitive quantum sensors.

*Data Availability Statement*
The data that support the findings of this study are available from the corresponding author upon reasonable request.

Acknowledgements:
J. Dietz acknowledges support from National Science Foundation RAISE TAQS grant number 1839164-PHY and support from the STC Center for Integrated Quantum Materials under National Science Foundation Grant No. DMR – 1231319. This work was performed in part at the Harvard University Center for Nanoscale Systems (CNS), a member of the National Nanotechnology Coordinated Infrastructure Network (NNCI), which is supported by the National Science Foundation under NSF ECCS award No. 1541959. The authors would like to further acknowledge Dr. Xingyu Zhang for fabrication support and Aaron Day for assistance with measurements.